\documentclass[aps,prc,preprint,superscriptaddress]{revtex4-1}

\usepackage{graphics}
\usepackage{dcolumn}
\usepackage{bm}
\usepackage{floatrow}
\usepackage{booktabs}
\usepackage{multirow}
\usepackage{siunitx}
\usepackage{longtable}

\begin{document}


\title{$^{54}$Fe($d$,$p$)$^{55}$Fe and the evolution of single neutron energies in the $N=29$ isotones}



\author{L. A. Riley}
\affiliation{Department of Physics and Astronomy, Ursinus College,
  Collegeville, PA 19426, USA}

\author{I.C.S. Hay} \affiliation{Department of Physics, Florida State University, Tallahassee, FL  32306, USA}

\author{L. T. Baby} \affiliation{Department of Physics, Florida
  State University, Tallahassee, FL 32306, USA}

\author{A. L. Conley} \affiliation{Department of Physics, Florida
  State University, Tallahassee, FL 32306, USA}

\author{P. D. Cottle} \affiliation{Department of Physics, Florida
  State University, Tallahassee, FL 32306, USA}

\author{J. Esparza} \affiliation{Department of Physics, Florida
  State University, Tallahassee, FL 32306, USA}

\author{K. Hanselman} \affiliation{Department of Physics, Florida
  State University, Tallahassee, FL 32306, USA}

\author{B. Kelly} \affiliation{Department of Physics, Florida
  State University, Tallahassee, FL 32306, USA}

\author{K. W. Kemper} \affiliation{Department of Physics, Florida
  State University, Tallahassee, FL 32306, USA}

\author{K. T. Macon} \affiliation{Department of Physics and Astronomy, 
Louisiana State University, Baton Rouge, Louisiana 70803, USA}

\author{G. W. McCann} \affiliation{Department of Physics, Florida
  State University, Tallahassee, FL 32306, USA}

\author{M.W. Quirin} \affiliation{Department of Physics and Astronomy, Ursinus College,
  Collegeville, PA 19426, USA}

\author{R. Renom} \affiliation{Department of Physics, Florida
  State University, Tallahassee, FL 32306, USA}

\author{R.L. Saunders} \affiliation{Department of Physics and Astronomy, Ursinus College,
  Collegeville, PA 19426, USA}

\author{M. Spieker} \affiliation{Department of Physics, Florida
  State University, Tallahassee, FL 32306, USA}

\author{I. Wiedenh\"over} \affiliation{Department of Physics, Florida
  State University, Tallahassee, FL 32306, USA}

\date{\today}

\begin{abstract}

A measurement of the $^{54}$Fe($d$,$p$)$^{55}$Fe reaction at 16 MeV was performed
using the Florida State University Super-Enge Split-Pole Spectrograph to determine single-neutron energies for the $2p_{3/2}$, $2p_{1/2}$, $1f_{5/2}$, $1g_{9/2}$ and $2d_{5/2}$ orbits.
Two states were observed that had not been observed in previous (d, p) measurements. In addition, we made angular momentum transfer, \textit{L}, assignments to four states and changed \textit{L} assignments from previous ($d$, $p$) measurements for nine more states.
The spin-orbit splitting between the $2p_{3/2}$ and $2p_{1/2}$ orbits is similar to that in the other $N=29$ isotones and not close to zero as a previous measurement suggested. While the $1f_{5/2}$ single neutron energy is significantly lower in $^{55}$Fe than in $^{51}$Ti, as predicted by a covariant density functional theory calculation, the single-neutron energy for this orbit in $^{55}$Fe is more than 1 MeV higher than the calculation suggests, although it is only 400 keV above the $2p_{1/2}$ orbit.  The summed spectroscopic strength we observed for the $1g_{9/2}$ orbit up to the single-neutron separation energy of 9.3 MeV is only 0.3.  This is surprising because the $1g_{9/2}$ orbit is predicted by Togashi \textit{et al.} to be located only 5.5 MeV above the $2p_{3/2}$ orbit. 
\end{abstract}


\maketitle


\section{Introduction}

The determination of single-nucleon energies is of central importance for understanding nuclear structure.  Furthermore, 
single-nucleon energies evolve as proton and neutron numbers change, and tracing this evolution is critical for understanding the nucleon-nucleon interactions taking place.

The island of inversion (IOI) centered on the $N=40$ nucleus $^{64}$Cr provides a particularly relevant example (for example, see Ref. \cite{Ga21}).  Na\"ively, the $N=40$ subshell gap between the $fp$ orbits and the $1g_{9/2}$ orbit should make the $N=40$ isotones semi-magic nuclei.  However, the data on these isotones reveal a much different situation, with deformed ground states and complex shape coexistence.  

There are intriguing questions regarding the $fp$ orbits as well.  In the $N=29$ isotope $^{55}$Fe, the previous measurement of the $^{54}$Fe($d$,$p$)$^{55}$Fe reaction that covered the entire range of excitation energies up to the 
neutron-separation energy of 9.3 MeV \cite{Fu63} suggested that the spin-orbit splitting between the $2p_{3/2}$ and 
$2p_{1/2}$ neutron orbits had been reduced to close to zero in this nucleus.  A more recent measurement of the same reaction \cite{Ma09} only covered excitation energies up to 4.5 MeV.    

In this article, we report on the determination of single-neutron energies in $^{55}$Fe using the $^{54}$Fe($d$,$p$)$^{55}$Fe reaction at 16 MeV.  We measured angular distributions for 38 states, the highest of which is at an excitation energy of 8.8 MeV.
Two states were observed that had not been observed in previous (d, p) measurements. In addition, we made angular momentum transfer, \textit{L}, assignments to four states and changed \textit{L} assignments from previous ($d$, $p$) measurements for nine more states.
We determined single-neutron energies for the $2p_{3/2}$, $2p_{1/2}$, $1f_{5/2}$, $1g_{9/2}$ and $2d_{5/2}$ orbits.  The total spectroscopic strength we observe for the $1g_{9/2}$ orbit is only 0.3, which is surprising because Togashi \textit{et al.} \cite{To15} predict that the $1g_{9/2}$ orbit is only 5.5 MeV above the $2p_{3/2}$ while we are able to measure states up to the single neutron separation energy of 9.3 MeV.  In addition, our results show that the spin-orbit splitting between the $2p_{1/2}$ and $2p_{3/2}$ neutron orbits is comparable to those found in other $N=29$ isotones.  Finally, we compare the present results for single-neutron energies as well as the corresponding information from other odd-A $N=29$ isotones to the results of a calculation using covariant density functional theory.  Our experimental result for the $1f_{5/2}$ single neutron energy in $^{55}$Fe is more than 1 MeV higher than the calculation.

\section{Experimental details and results}

A deuteron beam, produced by a SNICS (source of negative ions by cesium sputtering) source with a deuterated titanium cone, was accelerated to an energy of 16 MeV by the 9 MV Super FN Tandem Van de Graaff Accelerator at the John D. Fox Laboratory at Florida State University. The beam was delivered to a Fe target of thickness 0.44 mg/cm$^2$ enriched to $95\%$ in \textsuperscript{54}Fe that was mounted in the target chamber of the Super-Enge Split-Pole Spectrograph. The spectrograph, which accepted a solid angle of 4.6 msr, was rotated from scattering angles of 15$^\circ$ to 50$^\circ$ at increments of 5$^\circ$ to measure angular distributions of protons from the  \textsuperscript{54}Fe($d,p$)\textsuperscript{55}Fe reaction.  Further details of the experimental setup are described in Ref. \cite{Ri21}. 



\begin{figure*}[th]
  \begin{center}
    \scalebox{1.2}{
      \includegraphics{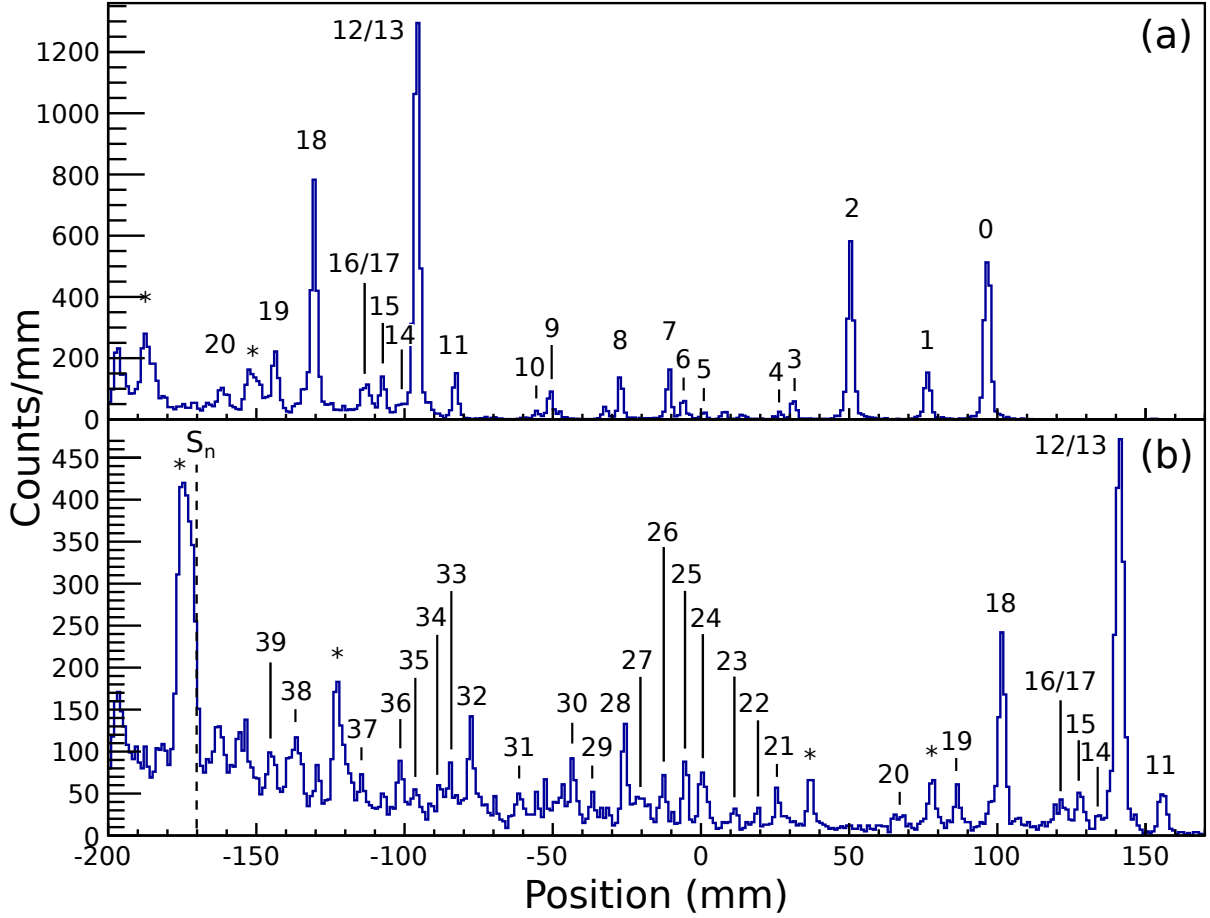}
    }
    \caption{\label{fig:spectrum} (Color online) Proton momentum spectrum at a laboratory angle of 30$^\circ$. Peaks corresponding to states of $^{55}$Fe are labeled. Peaks labeled with asterisks are contaminants. The spectrum is shown as a function of position in the focal plane detector.  } 
  \end{center}
\end{figure*}

A representative proton magnetic rigidity spectrum collected at a scattering angle of 30$^\circ$ is shown in Fig.~\ref{fig:spectrum}.

\begin{center}
\begin{longtable}{cccccccc}
  \caption{\label{tab:states} Excitation energies, angular-momentum transfer, and $J^\pi$ assignments, single-neutron orbits used for the \textsc{fresco} analysis, and the spectroscopic factors for states of \textsuperscript{55}Fe populated in the present work. Energies for states are taken from:  a = \cite{NNDC55}, b = \cite{Ma09}, and c = present work.  Established $J^\pi$ assignments are from Ref.~\cite{NNDC55}. Tentative $J^\pi$ assignments based on $L$ values determined in the present work are discussed in the text. When more than one possible orbit is given for a state, the spectroscopic factors assuming both orbits are shown.}\\

\hline

Label & $E_x$ (keV) & $E_x$ Ref. & \textit{L} & J$^\pi$ & orbit & S & Comments \\ 
\hline
\endfirsthead

\multicolumn{7}{c}

{{\bfseries \tablename\ \thetable{} --- continued from previous page}} \\
\hline
Label & $E_x$ (keV) & E Ref. & \textit{L} & J$^\pi$ & orbit & S & Comments \\
\hline
\endhead

\hline \multicolumn{7}{r}{{Continued on next page}} \\ \hline
\endfoot

\hline\hline
\endlastfoot

        0  & 0           & a & 1 & $\frac{3}{2}^-$ & $2p_{3/2}$ & 0.34(5) & \\
        1  & 411.4(2)  & a & 1 & $\frac{1}{2}^-$ & $2p_{1/2}$ & 0.22(3) & \\
        2  & 931.3(1) & a & 3 & $\frac{5}{2}^-$ & $1f_{5/2}$ & 0.30(5) & \\
        3  & 1316.5(1)  & a & 3  & $\frac{7}{2}^-$ & $1f_{7/2}$  & 0.017(3) &   \\
        4  & 1408.4(1)  & a & 3  & $\frac{7}{2}^-$ & $1f_{7/2}$ & 0.007(1) &  \\
        5  & 1918.3(5)  & a & 1 & $\frac{1}{2}^-$ & $2p_{1/2}$ & 0.032(5) &  \\
        6  & 2051.7(4)  & a & 1 & $\frac{3}{2}^-$ & $2p_{3/2}$ & 0.045(7) & \\
        7  & 2144.0(3)  & a & 3 & $\frac{5}{2}^-$ & $1f_{5/2}$ & 0.077(12) & \\
        8  & 2470.2(6)  & a & 1 & $\frac{3}{2}^-$ & $2p_{3/2}$ & 0.076(11) & \\
        9  & 2938.9(4)  & a & 3 & $\frac{7}{2}^-$ & $1f_{7/2}$ & 0.023(3) & \\
        10 & 3028.5(7)  & a & 1 & $\frac{3}{2}^-$ & $2p_{3/2}$ & 0.0084(12) & \\
        11 & 3552.3(8)  & a & 1 & $\frac{3}{2}^-$ & $2p_{3/2}$ & 0.064(10) & \\
        12 & 3790.3(8)  & a & 1 & $\frac{1}{2}^-$ & $2p_{1/2}$ & 0.33(5) & \\
        13 & 3804(2) & b & 4 & $\frac{9}{2}^+$ & $1g_{9/2}$ & 0.28(4) & \\
        14 & 3906.7(8)  & a & 3 & $\frac{5}{2}^-$ & $1f_{5/2}$ & 0.028(4) & Refs. \cite{Ma09,NNDC55} have $L=1$ \\
        15 & 4057(10)  & a & 3 & $\frac{5}{2}^-$ & $1f_{5/2}$ & 0.030(5) & \\
        16 & 4117 &  & & & & & $2p_{3/2}$ and $S=.0137(2)$ from \cite{Ma09} \\
        17 & 4134 &  & & & & & $1f_{5/2}$ and $S=.0066(3)$ from \cite{Ma09} \\
        18 & 4463(10)  & a & 2 & $\frac{5}{2}^+$ & $2d_{5/2}$ & 0.056(8) \\
        19 & 4708.3(7)  & a & 2 & $\frac{5}{2}^+$ & $2d_{5/2}$ & 0.019(3) \\
        20 & 5118(3)  & a & 1 & $\frac{1}{2}^-$ & $2p_{1/2}$ & 0.039(6) \\
        21 & 5839(10)  & a & 3 & $\frac{5}{2}^-$ & $1f_{5/2}$ & 0.046(7) & $L$ not previously measured \\
             &                 &    & 2 & $\frac{5}{2}^+$ & $2d_{5/2}$ & 0.012(2) & May be either $L=2$ or $L=3$ \\
        22 & 5955(10) & a & 2 & $\frac{5}{2}^+$ & $2d_{5/2}$ & 0.0091(14) & Ref. \cite{NNDC55} has $L=(0)$ \\
        23 & 6059(10)  & a & 3 & $\frac{5}{2}^-$ & $1f_{5/2}$ & 0.026(4) & Ref. \cite{NNDC55} has $L=2$ \\
        24 & 6282(10)  & a & 3 & $\frac{5}{2}^-$ & $1f_{5/2}$ & 0.086(13) & Ref. \cite{NNDC55} has $L=0$ \\
        25 & 6374(10)  & a & 2 & $\frac{5}{2}^+$ & $2d_{5/2}$ & 0.029(4) & $L$ not previously measured\\
        26 & 6495(10) & a  & 3 & $\frac{5}{2}^-$  & $1f_{5/2}$ & 0.065(10) & Ref. \cite{NNDC55} has $L=2$\\
        27 & 6628(10) & a  & 2 & $\frac{5}{2}^+$ & $2d_{5/2}$ & 0.015(2) & \\
        28 & 6776(10) & a  & 2 & $\frac{5}{2}^+$ & $2d_{5/2}$ & 0.038(6) & \\
        29 & 6916(10) & a  & 2 & $\frac{5}{2}^+$ & $2d_{5/2}$ & 0.019(3) & \\
        30 & 7030(10) & a  & 2 & $\frac{5}{2}^+$ & $2d_{5/2}$ & 0.025(4) & $L$ not previously measured \\
        31 & 7369(10) & a & 4 & $\frac{9}{2}^+$ & $1g_{9/2}$ & 0.020(3) & Ref. \cite{NNDC55} has $L=2$ \\
        32 & 7614(10) & a & 2 & $\frac{5}{2}^+$ & $2d_{5/2}$ & 0.027(4) & \\
        33 & 7762(10) & c  & 3 & $\frac{5}{2}^-$ & $1f_{5/2}$ & 0.046(7) & Not previously observed via ($d$,$p$)\\
        34 & 7808(10) & a & 1 & $\frac{1}{2}^-$ & $2p_{1/2}$ & 0.054(8) & Ref. \cite{NNDC55} has $L=2+(0)$ \\
        35 & 7938(10) & a & 1 & $\frac{1}{2}^-$ & $2p_{1/2}$ & 0.047(7) & $L$ not previously measured \\
             &                &     & 4 & $\frac{9}{2}^+$ & $1g_{9/2}$ & 0.017(3) & May be either $L=1$ or $L=4$ \\
        36 & 8028(10) & a & 2 & $\frac{5}{2}^+$ & $2d_{5/2}$ & 0.017(3) &  \\
        37 & 8264(10) & a & 3 & $\frac{5}{2}^-$ & $1f_{5/2}$ & 0.032(5) & Ref. \cite{NNDC55} has $L=2$ \\
        38 & 8660(10) & c & 2 & $\frac{5}{2}^+$ & $2d_{5/2}$ & 0.025(4) & State not previously observed \\
       39 & 8843(10) & a & 2 & $\frac{5}{2}^+$ & $2d_{5/2}$ & 0.014(2) & Ref. \cite{NNDC55} has $L=0$ \\

\end{longtable}
\end{center}

\begin{figure*}[h]
  \begin{center}
    \scalebox{0.6}{
      \includegraphics{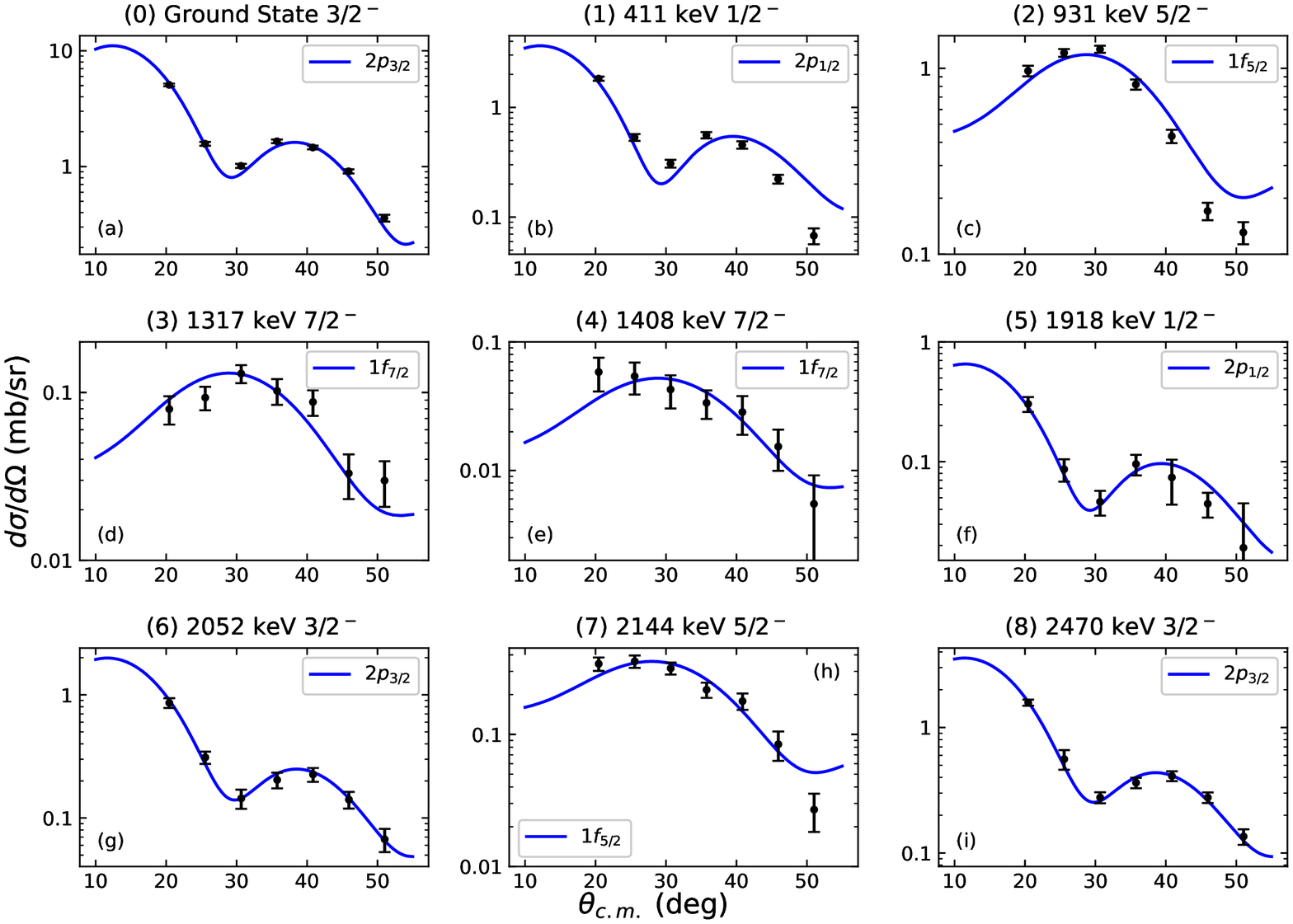}
    }
    \caption{\label{fig:angdist1} (Color online) Measured proton angular distributions from the $^{54}$Fe$(d,p)^{55}$Fe reaction compared with \textsc{fresco} calculations described in the text.  Panels (a) to (i) correspond to the states 0-8 in Table~\ref{tab:states}.}
    \end{center}
\end{figure*}

\begin{figure*}[h]
  \begin{center}
    \scalebox{0.6}{
      \includegraphics{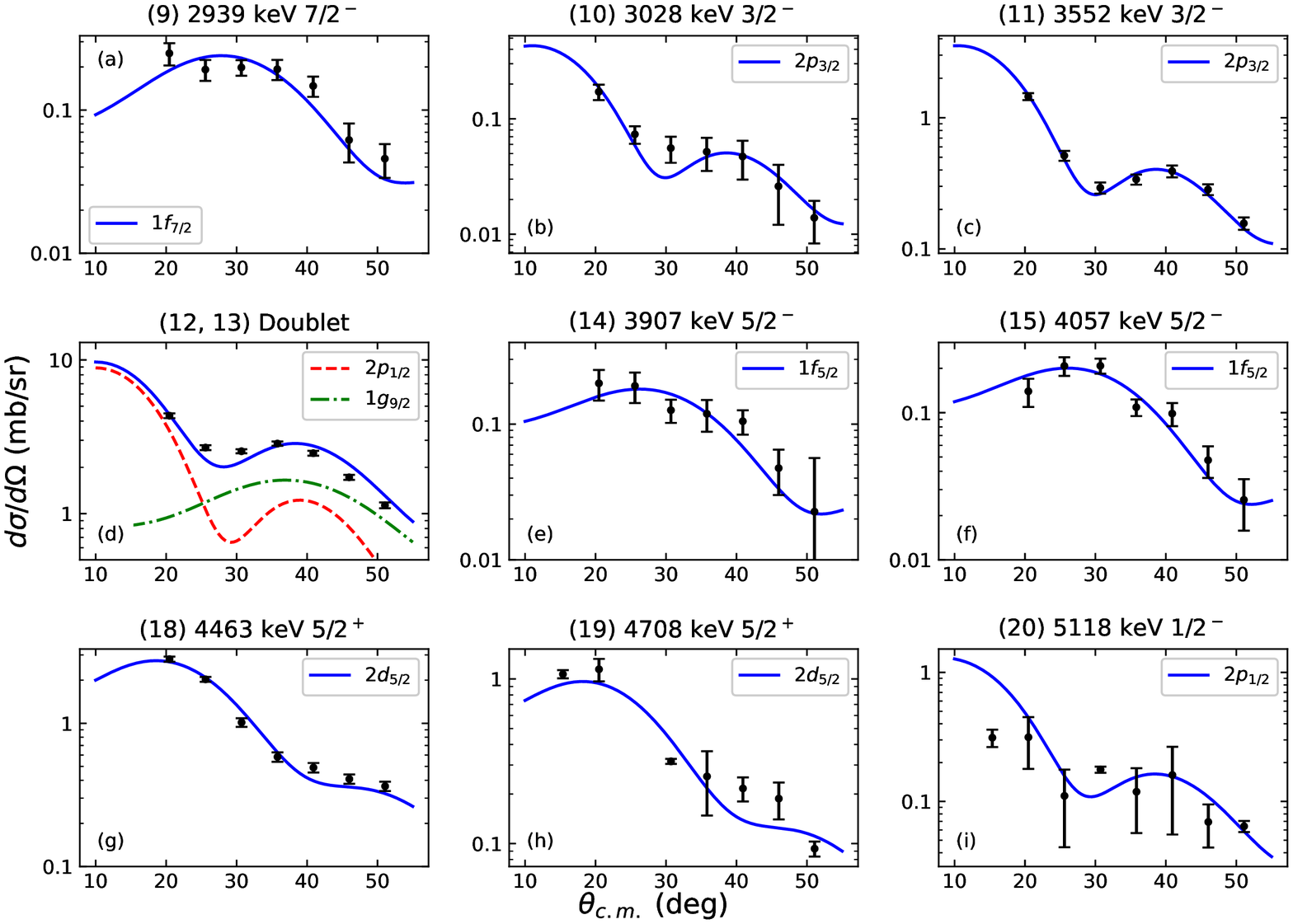}
    }
    \caption{\label{fig:angdist2} (Color online) Measured proton angular distributions from the $^{54}$Fe$(d,p)^{55}$Fe reaction compared with \textsc{fresco} calculations described in the text.   Panels (a) to (i) correspond to the states 9-15 and 18-20 in Table~\ref{tab:states}.}
    \end{center}
\end{figure*}

\begin{figure*}[h]
  \begin{center}
    \scalebox{0.6}{
      \includegraphics{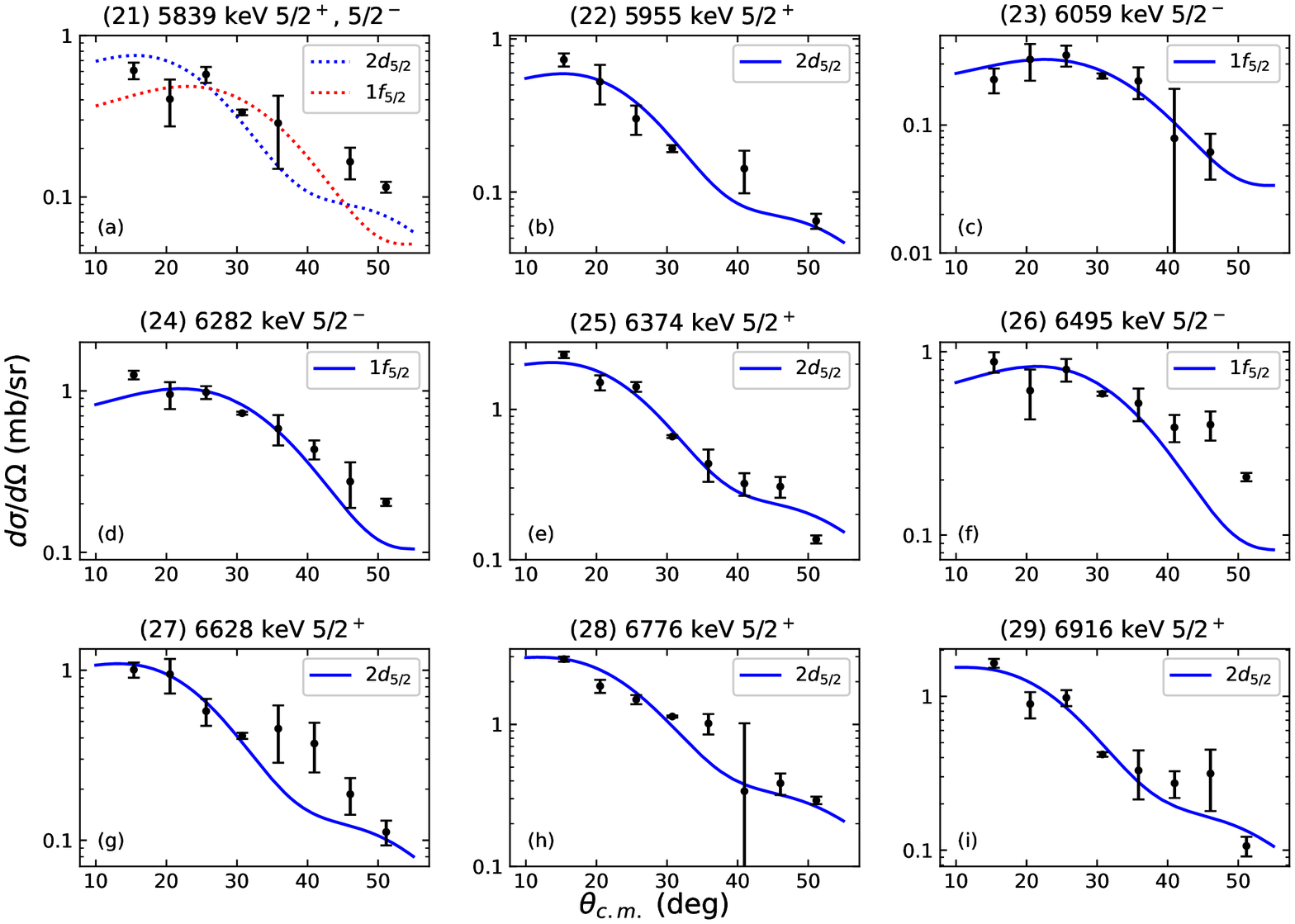}
    }
    \caption{\label{fig:angdist3} (Color online) Measured proton angular distributions from the $^{54}$Fe$(d,p)^{55}$Fe reaction compared with \textsc{fresco} calculations described in the text.   Panels (a) to (i) correspond to the states 21-29 in Table~\ref{tab:states}.}
    \end{center}
\end{figure*}

\begin{figure*}[h]
  \begin{center}
    \scalebox{0.6}{
      \includegraphics{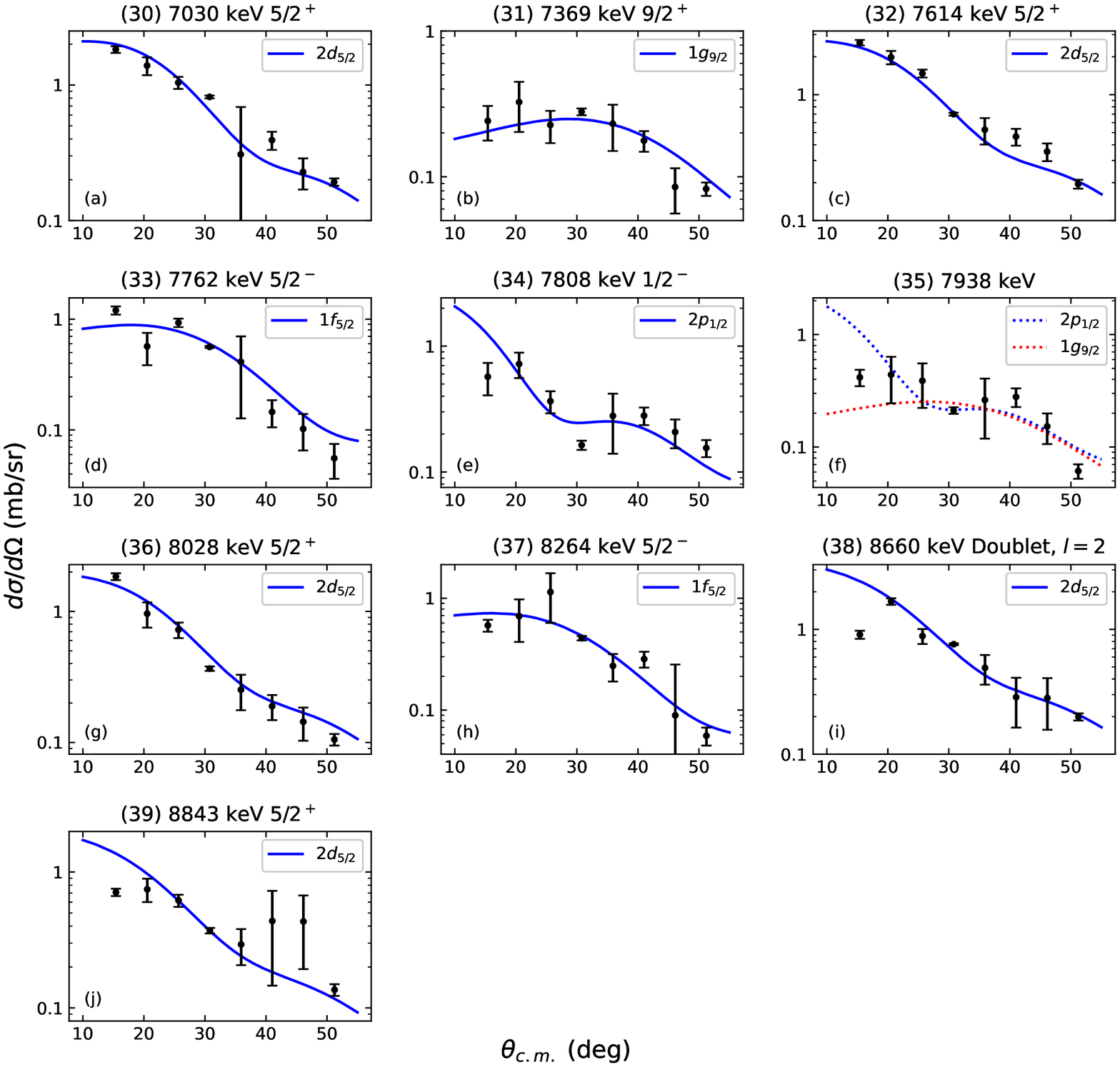}
    }
    \caption{\label{fig:angdist4} (Color online) Measured proton angular distributions from the $^{54}$Fe$(d,p)^{55}$Fe reaction compared with \textsc{fresco} calculations described in the text.   Panels (a) to (j) correspond to the states 30-39 in Table~\ref{tab:states}.}
    \end{center}
\end{figure*}

\begin{table*}
  \caption{\label{tab:omps} Optical potential parameters used in \textsc{fresco} calculations in the present work determined using Refs. \cite{Joh70} and \cite{Wal76} as described in the text.}
  \begin{tabular}{ccccccccccccccc}\hline\hline
             & $V_V$ & $r_V$ & $a_V$& $W_V$ & $r_W$ & $a_W$ & $W_D$ & $r_D$ & $a_D$ & $V_{so}$ & $W_{so}$ & $r_{so}$ & $a_{so}$ & $r_C$  \\
             & (MeV) & (fm) & (fm)  & (MeV) & (fm)  & (fm)  & (MeV)& (fm)  & (fm)  & (MeV)   & (MeV)    & (fm)     & (fm)    & (fm)   \\\hline
    d+$^{54}$Fe & 104.5 & 1.20 & 0.702 & 1.22  & 1.20  & 0.702 & 15.0 & 1.28  & 0.584 & 11.3    & -0.012   & 1.01     & 0.621   & 1.26   \\
    p+$^{55}$Fe &  53.1 & 1.20 & 0.670 & 1.28  & 1.20  & 0.670 & 8.15 & 1.28  & 0.547 & 5.54    & -0.067   & 1.02     & 0.590   & 1.26   \\\hline\hline
  \end{tabular}
\end{table*}

The magnetic rigidity spectrum measured at each scattering angle was fit using a linear combination of Gaussian functions with a quadratic background. The proton yields corresponding to each state in \textsuperscript{55}Fe were used to produce the measured proton angular distributions shown in Figs.~\ref{fig:angdist1}-\ref{fig:angdist4}.  The absolute cross sections were determined to be accurate to an uncertainty of $15\%$, with contributions from uncertainties in charge integration, target thickness and solid angle. 

To extract spectroscopic factors from the present angular distributions, calculations that use the adiabatic approach for generating the entrance channel deuteron optical potentials (as developed by Johnson and Soper \cite{Joh70}) were used.  The potential was produced using the formulation of Wales and Johnson \cite{Wal76}.  Its use takes into account the possibility of deuteron breakup and has been shown to provide a more consistent analysis as a function of bombarding energy \cite{De05} as well as across a large number of ($d,p$) and ($p,d$) transfer reactions on $Z=3-24$ target nuclei \cite{Le07}.  The proton-neutron and neutron-nucleus global optical potential parameters of Koning and Delaroche \cite{Kon03} were used to produce the deuteron potential as well as the proton-nucleus optical potential parameters needed for the exit channel of the ($d,p$) transfer calculations, in keeping with the nomenclature of Ref. \cite{De05}.  The angular momentum transfer and spectroscopic factors found in Table~\ref{tab:states} were determined by scaling these calculations, made with the \textsc{fresco} code \cite{Tho88}, to the proton angular distributions. Optical potential parameters are listed in Table~\ref{tab:omps}.  The overlaps between 
$^{55}$Fe and $^{54}$Fe$+n$ were calculated using binding potentials of Woods-Saxon form whose depth was varied to reproduce the given state's binding energy with geometry parameters of $r_0=1.25$ fm and $a_0=0.65$ fm and a Thomas spin-orbit term of strength $V_{so}=6$ MeV that was not varied.


We were able to associate 37 of the 38 states we measured with states listed in the most recent evaluation of data on 
$^{55}$Fe \cite{NNDC55}.  Of those 37, 36 were observed via the ($d$,$p$) reaction by Fulmer and McCarthy \cite{Fu63}.  The state we observed at 7762 keV is likely that observed via the $^{56}$Fe($p$,$d$)$^{55}$Fe reaction at 7780(50) keV.  

One of the states observed here was not readily identifiable with a state listed in Ref. \cite{NNDC55} but was observed in the $^{54}$Fe($d$,$p$)$^{55}$Fe study performed by Mahgoub \textit{et al.} \cite{Ma09}.  This state is the 
$\frac{9}{2}^+$ state at 3804(2) keV, which in our experiment formed a doublet with the 3790.3 keV $\frac{1}{2}^-$  state.  We were unable to resolve these two peaks in our experiment, but the Mahgoub experiment had sufficient resolution to do so.  To extract spectroscopic factors for the two states from our data, we performed a chi-square minimization procedure that fit the sum of two angular distributions --- one for $1g_{9/2}$ and the other for $2p_{1/2}$ --- to the data.  We allowed the spectroscopic factors for each of these two states to vary freely.  The two spectroscopic factors resulting from this fitting procedure, which are shown in Table I, were almost identical to those deduced by Mahgoub \textit{et al.}.  The fit for this compound peak is shown in Fig. 3. 

We observed a peak in the energy range 4110-4140 keV that has a width consistent with it being a complex of two or more states.  Once again, the experiment of Ref. \cite{Ma09} with its better energy resolution was able to resolve this complex into four individual peaks, two of which were contaminant peaks.  Since the present experiment is unable to resolve these four peaks, we adopt the $L$ transfer values and spectroscopic factors for states at 4117 and 4134 keV from Ref. \cite{Ma09}.


We observed a state at 8660 keV that had not previously been observed with any experimental probe. In fact, this peak is broad, suggesting that it is a doublet.  

Twelve of the 38 states we measured in this experiment have $L=2$ transfers, and for a thirteenth (5839 keV) we are unable to distringuish between $L=2$ and $L=3$.  It is most likely that this strength comes from the $2d_{5/2}$ orbit located above the $N=50$ major shell closure, although we are not aware of any theoretical explorations of the occurrence of $2d_{5/2}$ strength in the bound states of $^{55}$Fe or any other $N=29$ isotones.

We observe a significant amount of strength from the $1g_{9/2}$ orbit.  Nearly all of the observed $1g_{9/2}$ strength is located in the 3804 keV state [$S=0.28(4)$], with a small amount more located in the 7369 keV state [$S=0.020(3)$].  The total amount of $1g_{9/2}$ strength observed here ($S=0.30$) is much smaller than the total observed strengths for the three negative parity orbits of interest here ($S=0.55$ for $2p_{3/2}$, 0.70 for $2p_{1/2}$ and 0.75 for $1f_{5/2}$).

Of course, the $L$ value for a state does not completely determine the state's $J^{\pi}$ value.  The study of Mahgoub \textit{et al.} \cite{Ma09} used a polarized beam to measure analyzing powers, which allowed them to make $J^{\pi}$ assignments for the states they observed.  However, they only measured states up to 4.5 MeV.  We have assumed that all $L=1$ states above 5 MeV have $J^{\pi}=1/2^-$ since in other $N=29$ isotones ($^{49}$Ca, $^{51}$Ti and $^{53}$Cr) the $1p_{1/2}$ neutron orbit is between 1.5 and 2.0 MeV above the $2p_{3/2}$ orbit \cite{Ri21}.  In addition, we assume that all $L=3$ states above 5 MeV have $J^{\pi}=5/2^-$ because the $1f_{7/2}$ neutron orbit is pushed by the spin-orbit interaction into the next lower major shell.  Finally, we assume that all of the states populated via $L=2$ transfer have $J^{\pi}=5/2^+$ since the $2d_{3/2}$ orbit is significantly higher in energy than the $2d_{5/2}$ orbit.


\section{Single-neutron energies in $^{55}$Fe}

The ($d$,$p$) reaction provides an opportunity to identify significant fragments of single neutron strength so that a single neutron energy can be determined by calculating the centroid of the observed fragments. 

The largest concentration of $2p_{3/2}$ strength is located in the ground state.  However, there are significant concentrations of $2p_{3/2}$ strength in the 2052, 2470, 3029, 3552 and 4117 keV states.  The spectroscopic factors for these six states yield a centroid of 1080(110) keV above the ground state energy.  The experimental uncertainty is calculated by taking into account the 15$\%$ uncertainties in the spectroscopic factors.

The first excited state at 411 keV has a significant amount of $2p_{1/2}$ strength [$S=0.22(3)$], but the 3790 keV state has an even larger concentration of $2p_{1/2}$ strength [$S=0.33(5)$].  There are other $2p_{1/2}$ fragments in the states at 1918, 5118 and 7808 keV.  In addition, we are unable to determine whether the 7938 keV state is populated through $L=1$ transfer (giving $J^{\pi}=\frac{1}{2}^-$)  or $L=4$ transfer (giving $J^{\pi}=\frac{9}{2}^+$).  This uncertainty in the assignment of the 7938 keV state is accounted for in calculating the experimental uncertainty of the $2p_{1/2}$ centroid.  Altogether, we find a $2p_{1/2}$ centroid of 3170(220) keV above the ground-state energy.  This gives an energy difference of 2090(250) keV between the single-neutron energies of the $2p_{1/2}$ and $2p_{3/2}$ orbits. 

The second excited state at 931 keV has the largest concentration of $1f_{5/2}$ strength, but there are nine more states in which there is a significant amount of $1f_{5/2}$ strength --- at 2144, 3907, 4057, 4134, 6059, 6282, 6495, 7762 and 8264 keV.  In addition, the state at 5839 keV could be populated through either $L=3$ (giving $J^{\pi}=\frac{5}{2}^-$) or $L=2$ (giving 
$J^{\pi}=\frac{5}{2}^+$).  The uncertainty in this assignment is taken into account in calculating the experimental uncertainty in the centroid, giving a centroid energy of 3580(180) keV above the ground-state energy, or 2500(250) keV above the $2p_{3/2}$ single-neutron energy. 

In addition to the negative parity neutron orbits, two positive parity orbits are evident in the spectrum.  There is a significant concentration [$S=0.28(4)$] of $1g_{9/2}$ strength in the 3804 keV state, and a smaller concentration [$S=0.020(3)$] in the 7369 keV state.  In addition, we are unable to determine whether the angular distribution for the 7938 keV state corresponds to an $L=1$ or $L=4$ transfer.  If it is an $L=4$ transfer --- which implies $1g_{9/2}$ --- then the spectroscopic factor for this state is 0.017(3).  This uncertainty contributes to the uncertainty in the $1g_{9/2}$ single neutron energy, which is 4150(110) keV above the ground state, or 3070(160) keV above the $2p_{3/2}$ single-neutron energy.  A caution regarding this result is in order because the sum of the spectroscopic factors of the states observed here is only 0.3 --- it is possible that the present experiment has missed some higher-lying small fragments of $g_{9/2}$ strength and that the true single-neutron energy for the $1g_{9/2}$ orbit is higher.

While the $1g_{9/2}$ orbit is located above the $N=40$ subshell closure, the $2d_{5/2}$ orbit is located in the next major shell, above the $N=50$ subshell closure.  Hence, it is striking that so many $2d_{5/2}$ fragments --- adding up to a total spectroscopic factor of 0.29 --- are observed in the present experiment.  There are at least twelve states (4463, 4708, 5955, 6374, 6628, 6776, 6916, 7030, 7614, 8028, 8660 and 8843 keV) populated via $L=2$ transfer.  As mentioned above, it is highly likely these are $J^{\pi}=5/2^+$ states because the $2d_{3/2}$ orbit is significantly higher in energy than the $2d_{5/2}$ orbit.  In addition, the 8660 keV state is likely a doublet (the spectroscopic factor in the table is determined by a fit to the angular distribution of the entire 8660 keV peak and therefore assumes that both states represented in the peak have $L=2$).  Furthermore, it is not clear from the angular distribution of the 5839 keV state whether it is populated via $L=2$ or $L=3$ transfer.  That uncertainty increases the uncertainty in the single-neutron energy we calculate.  Our result is that the 
$2d_{5/2}$ single-neutron energy is 6550(14) keV above the ground state or 5470(140) keV above the $2p_{3/2}$ orbit.  As in the case of the $1g_{9/2}$ orbit, a relatively small percentage of the expected $2d_{5/2}$ strength is observed here (the total of the observed spectroscopic factors is 0.3), so it is quite possible that the true single-neutron energy is significantly higher than 5.5 MeV.

\section{Discussion}

\begin{figure}[h]
  \begin{center}
    \scalebox{0.33}{
      \includegraphics{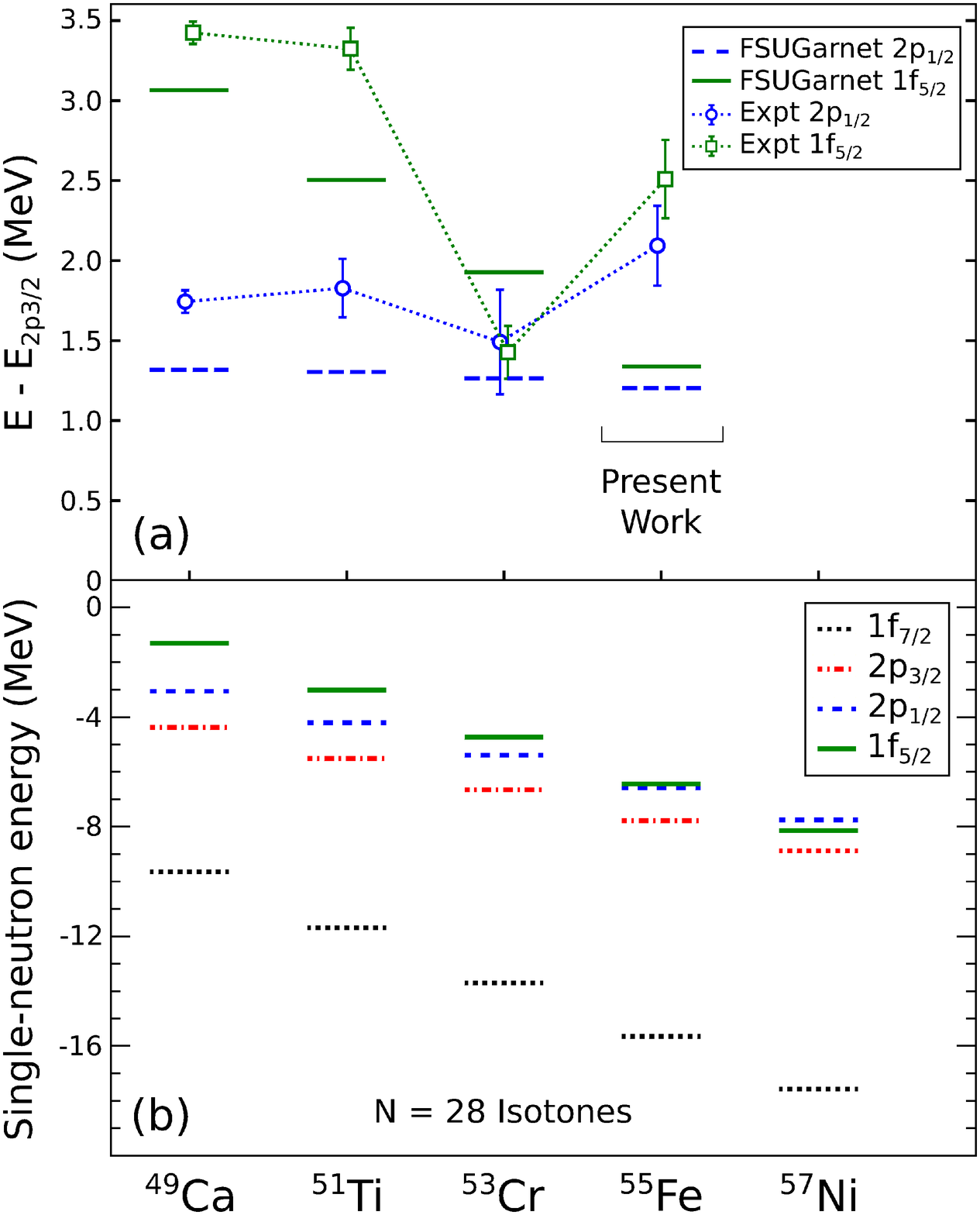}
    }
    \caption{\label{fig:sn_energies} (Color online) (a) Measured $1f_{5/2}$ and $2p_{1/2}$ single-neutron energy centroids, relative to the $2p_{3/2}$ energy, from the present work and Refs.~\cite{NNDC49,Ri21,NNDC53,NNDC55} compared with the covariant density functional theory approach described in the text. (b)  Single-neutron binding energies calculated using the covariant density functional theory.} 
    \end{center}
\end{figure}

We will first address an important but unsurprising result from the present experiment: the spin-orbit splitting between the $2p_{3/2}$ and $2p_{1/2}$ orbits is not zero, as the ($d$,$p$) results of Fulmer and McCarthy \cite{Fu63} (as compiled in \cite{NNDC55}) implied.  The observation of Maghoub \textit{et al.} \cite{Ma09} that there is a concentration of $2p_{1/2}$ strength at 3.8 MeV mostly settled the issue, but because that study only measured states up to an excitation energy of 4.5 MeV, their data is not sufficient to answer the question of the spin-orbit splitting definitively.  The present results do so.


Panel (a) of Fig. 6 shows the evolution of the $2p_{1/2}$-$2p_{3/2}$ and $1f_{5/2}$-$2p_{3/2}$ single neutron energy differences (or gaps) as a function of proton number for the $N=29$ isotones for which the ($d$,$p$) reaction can be measured using a stable target.  The results for $^{49}$Ca are extracted from the $^{48}$Ca($d$,$p$)$^{49}$Ca study of Ref. \cite{Uo94}.  That study used a polarized deuteron beam, so there is no uncertainty about $J^{\pi}$ values for the states observed. 

The $^{51}$Ti results are taken from the $^{50}$Ti($d$,$p$)$^{51}$Ti study reported in Ref. \cite{Ri21}.  That article includes a figure of single-neutron energies similar to Fig. 6; however, in extracting single neutron energies here we have assumed that all of the states above 4.5 MeV excitation energy populated in $L=1$ transfer have $J^{\pi}=\frac{1}{2}^-$ (which would make them $2p_{1/2}$ states) and all states in that energy range populated via $L=3$ transfer have $J^{\pi}=\frac{5}{2}^-$ (and are therefore $1f_{5/2}$ states).

The available data on $^{52}$Cr($d$,$p$)$^{53}$Cr \cite{NNDC53} were not taken with polarized deuteron beams, so there is significant uncertainty about the $J^{\pi}$ values for states populated in $L=1$ transfer at energies as low as 2.4 MeV.  So we will adopt the value of 1488(326) keV for the $2p_{1/2}$-$2p_{3/2}$ single neutron energy gap given in Ref. \cite{Ri21}.  We will also adopt the value of 1424(165) keV for the $1f_{5/2}$-$2p_{3/2}$ gap from Ref. \cite{Ri21}.   

Figure 6(a) also includes the results of calculations made in the framework of covariant density functional theory using the covariant energy density functional FSUGarnet\,\cite{che15} that was calibrated using the fitting protocol described 
in Ref.\,\cite{che14}.  These calculations were first published in \cite{Ri21} and the calculations are explained there.  Panel 6(b) shows the same calculated single-neutron energies as binding energies.  

The very first thing to notice about the experimental points in Fig. 6(a) is that the point for the $1f_{5/2}$ single neutron energy in $^{53}$Cr seems to be troublesome when compared to the results for the neighboring $^{51}$Ti and $^{55}$Fe isotones.  A remeasurement of the $^{52}$Cr($d$,$p$)$^{53}$Cr reaction is clearly called for, and such a remeasurement might also result in a revision of the $2p_{1/2}$ single neutron energy in that nucleus.  

Figure 6(a) also shows that the covariant density functional theory underestimates the energy of the $1f_{5/2}$ orbit.  The discrepancy starts off at about 300 keV in $^{49}$Ca, grows to 700 keV in $^{51}$Ti and then opens up to more than 1 MeV in $^{55}$Fe.  Furthermore, the calculation underestimates the $2p_{1/2}$-$2p_{3/2}$ spin-orbit splitting.  The calculated values of the spin-orbit splitting are less than 1.5 MeV for all four isotones shown.  In contrast, the experimental results for $^{49}$Ca, $^{51}$Ti and $^{55}$Fe are all larger than 1.5 MeV.

Nevertheless, both the calculated and experimental results for the $1f_{5/2}$ single neutron energies show a decrease in this energy as the proton number increases.  That trend can be understood in a straightforward way: as the $1f_{7/2}$ proton orbit fills, the attractive proton-neutron interaction pulls the single neutron energy of the $1f_{5/2}$ downward.  In $^{55}$Fe, the $1f_{5/2}$ orbit is only 400 keV above the $2p_{1/2}$ orbit.

It is not surprising that the covariant density functional theory calculation fits the shell gaps in $^{49}$Ca better than in the $N=29$ isotones with higher $Z$ values:  $^{48}$Ca was one of the nuclei used in the optimization procedure for the functional used here \cite{che14}.   

The $1g_{9/2}$ orbit is generally understood to be located well above the $fp$ orbits and above the $N=40$ subshell closure.  Togashi \textit{et al.} \cite{To15} calculated effective single neutron energies in the Fe isotopes, finding that in $^{55}$Fe the $1g_{9/2}$ orbit is located approximately 5.5 MeV above the $2p_{3/2}$ orbit and 2.5 MeV above the $2p_{1/2}$ orbit, which they calculated to be above the $1f_{5/2}$ orbit in this nucleus.  The centroid of the $1g_{9/2}$ strength we observe, 3.1 MeV above the $2p_{3/2}$ orbit, is considerably lower than the effective single neutron energy predicted by Togashi \textit{et al.}.  

However, perhaps the most important conclusion we can draw regarding the $1g_{9/2}$ orbit is that the sum of the spectroscopic factors we observed for this orbit was only 0.3.  The prediction of Togashi \textit{et al.} that the $1g_{9/2}$ orbit is 5.5 MeV above the $2p_{3/2}$ orbit suggests that we should see most of the $1g_{9/2}$ strength (that is, the sum of the spectroscopic factors of the observed states should be greater than 0.5)  in the present experiment since we are able to measure states up to the single neutron separation energy of 9.3 MeV.  Of course, we do not.  The observed
$1g_{9/2}$ strength observed in $^{55}$Fe is somewhat larger than that seen in $^{51}$Ti, which is 0.2 \cite{Ri21}.  However, the single neutron separation energy in $^{51}$Ti is only 6.4 MeV, so it is less surprising that there is so much missing $1g_{9/2}$ strength in that nucleus.  

Togashi \textit{et al.} also calculated the effective single neutron energy of the $2d_{5/2}$ orbit in $^{55}$Fe, finding that it is approximately 8 MeV above the $2p_{3/2}$ orbit.  Once again, our $2d_{5/2}$ centroid, 5.5 MeV above the $2p_{3/2}$ orbit, is significantly lower than the effective single neutron energy calculated by Togashi \textit{et al.}  As in the case of the $1g_{9/2}$, only about 30$\%$ of the expected strength was observed.  However, that is less surprising for the $2d_{5/2}$ orbit than it is for the $1g_{9/2}$ orbit because the former orbit is expected to be much higher in energy.

\section{Conclusions}

We performed a measurement of the $^{54}$Fe($d$,$p$)$^{55}$Fe reaction at 16 MeV using a Super-Enge Split-Pole Spectrograph.  Two states were observed that had not been observed in previous ($d$,$p$) measurements, and the \textit{L} transfer values for 13 previously measured states were either changed or measured for the first time.  We extracted 
single-neutron energies for the $2p_{3/2}$, $2p_{1/2}$, $1f_{5/2}$, $1g_{9/2}$ and $2d_{5/2}$ orbits.  Even though the prediction by Togashi \textit{et al.} suggests that we should be able to observe most of the $1g_{9/2}$ strength in the present experiment, the sum of the spectroscopic factors of the $1g_{9/2}$ seen here was only 0.3.  There is a substantial spin-orbit splitting between the $2p_{3/2}$ and $2p_{1/2}$ orbits.  In addition, the single-neutron energy of the $1f_{5/2}$ orbit appears to decline as the proton number increases, although the result for $^{53}$Cr is anomalous.  The decline of the 
$1f_{5/2}$ single neutron energy as the $1f_{7/2}$ proton orbit fills is expected because of the attractive spin-orbit interaction between the two orbits.  A remeasurement of the $^{52}$Cr($d$,$p$)$^{53}$Cr reaction should be performed. 

\begin{acknowledgments}
A target provided by the Center for Accelerator Target Science at Argonne National Laboratory was used in this work.  We thank J. Piekarewicz for the covariant density functional theory calculations.  This work was supported by the National Science Foundation through Grant No. PHY-2012522.
\end{acknowledgments}


%

\end{document}